\documentclass[
twocolumn,
amsmath,amssymb,
aps,
prb,
floatfix,
longbibliography,
noeprint,
]{revtex4-2}

\pdfoutput=1

\usepackage{xcolor,mathrsfs,dsfont}
\definecolor{darkblue}{RGB}{0,0,150}
\definecolor{nightblue}{RGB}{0,0,100}

\usepackage{graphicx,mathtools,bm}
\usepackage[
colorlinks,
citecolor=darkblue,
linkcolor=darkblue,
urlcolor=nightblue]{hyperref}

\usepackage[english]{babel}
\usepackage[babel,kerning=true,spacing=true]{microtype}

\newcommand{\bp}{\mathbf{p}}

\newcommand{\lam}{\lambda}
\newcommand{\alf}{\alpha}
\newcommand{\eps}{\epsilon}

\newcommand{\tr}{\text{tr}}

\newcommand{\cG}{\mathcal{G}}
\newcommand{\cA}{\mathcal{A}}

\newcommand{\sG}{\mathscr{G}}

\renewcommand{\Im}{\ensuremath{\mathrm{Im}\,}}
\renewcommand{\Re}{\ensuremath{\mathrm{Re}\,}}

\newcommand{\beq}{\begin{equation}}
\newcommand{\eeq}{\end{equation}}

\bibpunct{[}{]}{,}{n}{}{}
\makeatletter
\def\NAT@def@citea{\def\@citea{\NAT@separator}}
\makeatother

\begin{document}

\title{
Bound on resistivity in flat-band materials due to the quantum metric
}
\author{Johannes Mitscherling${}^1$}
\author{Tobias Holder${}^2$}
\affiliation{${}^1$Max Planck Institute for Solid State Research, Heisenbergstrasse
1, 70569, Stuttgart, Germany\\
${}^2$Department of Condensed Matter Physics, Weizmann Institute of Science, Rehovot, Israel 76100}
\date{\today}

\begin{abstract}
%\begin{linenumbers}
The quantum metric is a central quantity of band theory but has so far not been related to many response coefficients due to its nonclassical origin. 
However, within a newly developed Kubo formalism for fast relaxation, the decomposition of the dc electrical conductivity into both classical (intraband) and quantum (interband) contributions recently revealed that the interband part is proportional to the quantum metric. 
Here, we show that interband effects due to the quantum metric can be significantly enhanced and even dominate the conductivity for semimetals at charge neutrality and for systems with highly quenched bandwidth.
This is true in particular for topological flat-band materials of nonzero Chern number, where for intermediate relaxation rates an upper bound exists for the resistivity due to the common geometrical origin of quantum metric and Berry curvature. 
We suggest to search for these effects in highly tunable rhombohedral trilayer graphene flakes.
%\end{linenumbers}
\end{abstract}

\maketitle

\section{Introduction}

Recent experimental progress has made it possible to investigate new two-dimensional quantum materials with a highly quenched bandwidth, which exhibit a rich phase diagram including unconventional superconducting phases, correlated insulating states and a cascade of magnetic and nonmagnetic flavor-ordered states~\cite{Cao2018,Cao2018a,Lu2019,Yankowitz2019,Sharpe2019,Zondiner2020,Liu2020c,Chen2020,Park2021,Hao2021,Zhou2021,Zhou2021a,Wang2020,Zhang2020}.
While the most interesting phenomena in the ordered states are due to many-body effects, these findings have also ignited renewed interest in understanding the normal state of moiré-type and periodic lattice systems with vanishing Fermi velocity~\cite{Polshyn2019,Patel2019,Cao2020,Ghiotto2021}.
However, remarkably little is known about the effects of a quenched bandwidth on the electrical conductivity.
In this work, we systematically investigate the interplay between band structure and conductivity using a noninteracting Kubo approach which allows to fully identify and quantify the various contributing factors.

The two paradigmatic flat-band systems are
twisted bilayer graphene (TBG), a van-der-Waals material where the second layer is put askew by a small angle~\cite{Bistritzer2011,Santos2012}; and rhombohedral trilayer graphene (RTG), where three graphene monolayers are layered on top of each other without twist, with atomic positions in the sequence A-B-C~\cite{Koshino2009,Zhang2010,Koshino2010,Mak2010}. 
Here, we will only discuss the latter material, of which high-mobility devices exist, which makes it a viable platform to study band structure effects, and where the mobility can be controlled precisely.
For theory, superconductivity in flat bands presents a challenge, because a classic result predicts a vanishing superfluid stiffness for that case~\cite{Nelson1977}. In the meantime, several indications have been found that a quantity known as the quantum metric prevents the superfluid stiffness from going to zero~\cite{Peotta2015, Liang2017, Hu2019, Julku2020}. The quantum metric has previously been connected to the size of a maximally localized Wannier state~\cite{Marzari1997,  Marzari2012} and is crucial in the modern theory of polarization~\cite{Resta2011}. Only recently, its importance was noticed in other contexts as well~\cite{Gao2015,Kolodrubetz2017,Gao2019a,Holder2020,Ahn2021}. 

The quantum metric is a central quantity of band theory, comparable to the Berry curvature. Both form the quantum geometric tensor, which captures the geometrical properties of the manifold that is formed by the Bloch eigenstates~\cite{Provost1980}.
Nevertheless, due to its nonclassical origin the quantum metric has not yet been related to many response coefficients. 
In particular, the quantum metric has not been measured neither in the superconducting nor in the normal state of both TBG and RTG. 

\begin{figure}
    \centering
     \includegraphics[width=.75\columnwidth]{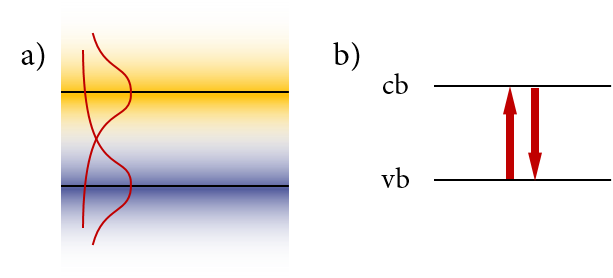}
    %\internallinenumbers
    \caption{
    Emergence of the longitudinal interband conductivity for a band structure with a finite broadening $\Gamma$. 
    (a) According to semiclassics, there is a finite density of states from the conduction band (cb) present at the valence band (vb), and vice versa, that is introduced by the band broadening in the respective other band. (b) According to perturbation theory the bands are defined sharply; instead a finite carrier lifetime allows virtual processes involving remote bands.}
    \label{fig:fig1}
\end{figure}

Recently, a new formulation of the Kubo formalism in two-band models for arbitrary band broadening $\Gamma$ (i.e., fast quasiparticle relaxation rate) was developed by one of us, which revealed that the static charge conductivity \emph{does} contain an interband contribution proportional to the quantum metric~\cite{Mitscherling2020}. However, in a metallic state the interband term is generically of order $\Gamma$, so that it is strongly suppressed compared to the intraband (Drude-type) response of order $\Gamma^{-1}$. 
The interband effect and its sensitivity to the relaxation rate can be understood either semiclassically as the overlap of separated bands upon band broadening; in perturbation theory one would instead attribute it to virtual excitations across the energy gap [cf. Fig.~\ref{fig:fig1}].
This is because at zero temperature the single-particle gap can only be traversed by placing some valence band states at energies across the gap. In the semiclassical view, this is the case after taking an ensemble average, while in the quantum perturbative picture, it is the case after averaging over time, due to transient (virtual) occupations. Similar reasoning has been successfully employed to explain the subgap response in the nonlinear optical conductivity~\cite{Kaplan2020} and for certain parts of the nonlinear dc conductivity~\cite{Holder2021a}.
In the following, we will establish several cases where interband effects due to the quantum metric can substantially contribute to or even dominate the conductivity.  

The key observation is that the conventional scaling only applies in the metallic state with a Fermi surface, where the quasiparticle velocities and the quantum metric are approximately constant. In contrast, consider for example a semimetal where the Fermi surface is just a point in momentum space, at which the quantum metric diverges. 
Another example is a flat band, where the vanishing bandwidth results in a very small Drude conductivity compared to the quantum metric contribution.
We note that these cases are in line with the interpretation of the quantum metric as a measure of the deformation of the wave packet as it traverses the lattice~\cite{Holder2021}, such deformations are facilitated by the large spread of a wave packet in real- or momentum space.

In the following, we show that the new Kubo formalism for fast relaxation is capable to treat these nonstandard cases. The formalism is based on a phenomenological quasiparticle relaxation rate $\Gamma$ of arbitrary size. This regularizes the spectral occupation, which results in smooth conductivity formulas and allow for a systematic expansion in $\Gamma$.
Since the previous formulation in Ref.~\cite{Mitscherling2020} was limited to two bands, we also present a generalization for multiband systems.
For semimetals at charge neutrality, we find that both quantum metric and intraband contribution become of same order in $\Gamma$. For example, the well-known result for the conductivity of a two-dimensional Dirac cone, $\sigma^{xx}=e^2/\pi h$~\cite{CastroNeto2009}, is constituted half by the intraband part and half by the quantum metric.
%For semimetals at charge neutrality, the divergence of the quantum metric is compensated by a vanishing spectral occupation. For a two-dimensional Dirac cone, we recover the well-known result $\sigma^{xx}=e^2/\pi h$~\cite{CastroNeto2009} independent of $\Gamma$. We find that both quantum metric and intraband contribution provide half of this result. For four different models, we find that the intraband and quantum metric contributions are of the same order in $\Gamma$.
%Flat band systems are particular interesting, since the vanishing quasiparticle velocity drastically reduces the intraband contribution. 
%We therefore expect the quantum metric to prominently enter the dc conductivity in systems with highly quenched bandwidth. 
%To this end, we calculate the normal state conductivity in the flat band limit for a band structure with arbitrary lifetime broadening. 
For systems with flat bands, as expected we find that the interband term can become larger than the Drude term. 
Here, the conductivity develops a plateau for an intermediate regime $W_\text{flat}<\Gamma<\Delta_\text{gap}$ of band broadening, with the lower limit of the plateau given by the small bandwidth $W_\text{flat}$ of the quenched band, while the band gap $\Delta_\text{gap}$ to neighboring bands presents the upper limit. 

We explain this finding as follows: The quantum metric $g$ and the Berry curvature $\Omega$ are the real and imaginary part of the quantum geometric tensor, respectively, which is positive semidefinite~\cite{Peotta2015}. This imposes a constraint on their components $g^{\alf\beta}$ and $\Omega^{\alf\beta}$, leading to a lower bound  $\int_{\bm{p}}\tr\ g\geq\int_{\bm{p}}|\Omega^{xy}|\geq  |C|/2\pi$ in two-dimensional systems~\cite{Roy2014, Peotta2015, Ozawa2021}, where $C$ is the Chern number. This property is also known as "Wannier obstruction"~\cite{Marzari1997, Matsuura2010, Marzari2012}. In two-dimensional flat-band systems in the regime $W_\text{flat}<\Gamma<\Delta_\text{gap}$ of band broadening, we find that the trace over the interband terms is proportional to $\int_{\bm{p}} \tr\ g$ and thus bounded from below. This implies a new upper bound on the resistivity, unrelated to previous suggestions for bounds due to strong interactions and the related fast thermalization of quasiparticles~\cite{Hartnoll2015,Lucas2017a,Hartman2017,Patel2019}. We emphasize that this bound is not universal and only holds within a certain regime of relaxation rates. 

We propose to use RTG devices of varying mobility as a platform to experimentally search for the quantum metric effect~\footnote{We note that the existence of high-mobility devices makes it very likely that in experiment the mobility can be tuned with relative ease.}. In particular, we suggest to study the temperature dependence of the conductivity and predict a conductivity minimum below room temperature as a result of decreasing classical (intraband) and increasing quantum metric (interband) contributions. In RTG, both intraband and interband contributions are of similar size over the full temperature range in contrast to perfect flat-band systems, where the energy scales are well separated. Nevertheless, we can clearly assign the conductivity minimum to the quantum metric, since a conductivity upturn due to thermal occupation of excited states takes place at much higher temperatures. 
%We find the minimum for a broad range of tunable parameters.

The paper is structured as follows: We present the Kubo formalism for fast relaxation for a general multiband system and present our key formulas for the conductivity. The role of band broadening is discussed for metals, semimetals, and flat bands, where the bound on the resistivity is derived. The theory is then applied to a generic flat-band model and RTG. We summarize our findings in the conclusions.

\section{Results}

\subsection{Conductivity formulas}

We consider a noninteracting quadratic tight-binding Hamiltonian for an $N$-band system
%
%\begin{linenomath*}
\begin{align}
    \label{eqn:H}
    H=\sum_\bp\Psi^\dag_\bp \lam^{}_\bp \Psi^{}_\bp \, ,
\end{align}
%\end{linenomath*}
%
where the spinor $\Psi_\bp=\big(c_{\bp,1},...\,,c_{\bp,N}\big)$ and its complex conjugate $\Psi^\dag_\bp$ involve the fermionic annihilation and creation operators of the $N$ bands at (crystal) momentum $\bp$, respectively. $\lam_\bp$ is the Hermitian $N\times N$ Bloch Hamiltonian matrix. We calculate the dc conductivity tensor $\sigma^{\alf\beta}$, which relates the spatial components of the current $j^\alf$ and the electric field $E^\beta$ via $j^\alf=\sum_\beta\sigma^{\alf\beta}E^\beta$. Generalizing the approach developed by one of us~\cite{Mitscherling2020} as presented in Appendix~\ref{sec:KuboFormalism}, we obtain the decomposition
%
%\begin{linenomath*}
\begin{align}
    \label{eqn:sigma}
    \sigma^{\alf\beta}=\sigma^{\alf\beta}_\text{intra}+\sigma^{\alf\beta,s}_\text{inter}+\sigma^{\alf\beta,a}_\text{inter}\, .
\end{align}
%\end{linenomath*}
%
The intraband contribution $\sigma^{\alf\beta}_\text{intra}=\sigma^{\beta\alf}_\text{intra}$ is symmetric, whereas the interband contribution has both a symmetric part $\sigma^{\alf\beta,s}_\text{inter}=\sigma^{\beta\alf,s}_\text{inter}$ and an antisymmetric part $\sigma^{\alf\beta,a}_\text{inter}=-\sigma^{\beta\alf,a}_\text{inter}$. The three contributions read
%
%\begin{linenomath*}
\begin{align}
    \label{eqn:Sintra}
    \sigma^{\alf\beta}_\text{intra}&\hspace{-1mm}=\hspace{-0.5mm}-\frac{e^2}{\hbar}\hspace{-1.5mm}\int\hspace{-2mm}\frac{d^d\bp}{(2\pi)^d}\hspace{-1.5mm}\int\hspace{-1.mm}d\eps\,\hspace{-0.5mm}f'(\eps)\,\hspace{-1mm}\sum_n\hspace{-0.5mm} w^{n}_{\bp,\text{intra}}(\eps)\,v^{\alf,n}_\bp v^{\beta,n}_\bp\!\!,\\
    \label{eqn:SinterS}
    \sigma^{\alf\beta,s}_\text{inter}&\hspace{-1mm}=\hspace{-0.5mm}-\frac{e^2}{\hbar}\hspace{-1.5mm}\int\hspace{-2mm}\frac{d^d\bp}{(2\pi)^d}\hspace{-1.5mm}\int\hspace{-1.mm}d\eps\,\hspace{-0.5mm}f'(\eps)\,\hspace{-1mm}\sum_{n,m}\hspace{-0.5mm} w^{nm,s}_{\bp,\text{inter}}(\eps)\,g^{\alf\beta,nm}_\bp\,,\\
    \label{eqn:SinterA}
    \sigma^{\alf\beta,a}_\text{inter}&\hspace{-1mm}=\hspace{-0.5mm}-\frac{e^2}{\hbar}\hspace{-1.5mm}\int\hspace{-2mm}\frac{d^d\bp}{(2\pi)^d}\hspace{-1.5mm}\int\hspace{-1.mm}d\eps\,\hspace{-0.5mm}f(\eps)\,\hspace{-0.5mm}\sum_{n,m}\hspace{-0.5mm} w^{nm,a}_{\bp,\text{inter}}(\eps)\,\Omega^{\alf\beta,nm}_\bp\,,
\end{align}
%\end{linenomath*}
%
where $f(\eps)=[\exp(\eps/k_B T)+1]^{-1}$ and $f'(\eps)$ are the Fermi distribution and its derivative, respectively. The momentum integration is performed over the $d$-dimensional Brillouin zone. 

We discuss the different quantities in the following. The intraband contribution \eqref{eqn:Sintra} involves a summation over all bands with their respective quasiparticle velocities $v^{\alf,n}_\bp=\partial_\alf E^n_\bp$, where $\partial_\alf$ denotes the momentum derivative in $\alf$ direction and $E^n_\bp$ is the $n$-th eigenvalue of $\lam_\bp$ with corresponding eigenvector $|n_\bp\rangle$. The two interband contributions \eqref{eqn:SinterS} and \eqref{eqn:SinterA} involve a summation over all pairs of bands with 
%
%\begin{linenomath*}
\begin{align}
    \label{eqn:gnm}
    &g^{\alf\beta,nm}_\bp=\Re\hspace{-1mm}\big[\cA^{\alf,nm}_\bp \cA^{\beta,mn}_\bp\big] \, ,\\[1mm]
    \label{eqn:Onm}
    &\Omega^{\alf\beta,nm}_\bp=-2\,\Im\hspace{-1mm}\big[\cA^{\alf,nm}_\bp \cA^{\beta,mn}_\bp\big] \, ,
\end{align}
%\end{linenomath*}
%
where $\cA^{\alf, nm}_\bp=i\langle n_\bp|\partial_\alf m_\bp\rangle$ is the Berry connection. We emphasize that the interband contribution can be interpreted as a pure quantum geometrical quantity. For a two-band system, note that \eqref{eqn:gnm} and \eqref{eqn:Onm} are the quantum metric and the Berry curvature, respectively, whereas a further trace over the bands is required to establish this connection for more than two bands [see \ref{sec:KuboFormalism}]. To highlight the geometric interpretation of the interband contributions, we will refer to $\sigma^{\alf\beta,s}_\text{inter}$ and $\sigma^{\alf\beta,a}_\text{inter}$ respectively as the {\it quantum metric} and the {\it Berry curvature contribution}. The three contributions to the conductivity \eqref{eqn:Sintra}-\eqref{eqn:SinterA} involve three spectral weighting factors
%
%\begin{linenomath*}
\begin{align}
    \label{eqn:wintra}
    w^{n}_{\bp,\text{intra}}(\eps)&=\pi\big[A^n_\bp(\eps)\big]^2\,,\\[1mm]
    \label{eqn:winterS}
    w^{nm,s}_{\bp,\text{inter}}(\eps)&=\pi(E^n_\bp-E^m_\bp)^2A^n_\bp(\eps)A^m_\bp(\eps) \, ,\\[1mm]
    \label{eqn:winterA}
    w^{nm,a}_{\bp,\text{inter}}(\eps)&=2\pi^2(E^n_\bp-E^m_\bp)^2\big[A^n_\bp(\eps)\big]^2A^m_\bp(\eps) \, ,
\end{align}
%\end{linenomath*}
%
which are particular combinations of the quasiparticle spectral functions
%
%\begin{linenomath*}
\begin{align}
    A^n_\bp(\eps)=\frac{1}{\pi}\frac{\Gamma}{\Gamma^2+(\eps+\mu-E^n_\bp)^2} \, ,
\end{align}
%\end{linenomath*}
%
where $\Gamma$ is the relaxation rate, which is assumed to be frequency- and momentum-independent as well as equal for all bands.
Note that $w^{nn,s}_{\bp,\text{inter}}(\eps)=w^{nn,a}_{\bp,\text{inter}}(\eps)=0$ for the diagonal components.

We repeat that no assumptions on the size of $\Gamma$ have been imposed to obtain the conductivity in Eq.~\eqref{eqn:sigma}; and its contributions as given by Eqs.~\eqref{eqn:Sintra}-\eqref{eqn:SinterA} hold for $\Gamma$ of arbitrary size. 
However, the results rely on the featureless form of the phenomenological relaxation rate $\Gamma$. A generalization, for instance, by using a frequency and momentum dependent $\Gamma(\bp,\omega)$ or a band dependent $\Gamma_{n}$ seems possible, but is beyond the scope of this work. Furthermore, we do not specify the physical origin of $\Gamma$, which could be due to interactions, impurity scattering, or coupling to the environment. Potential vertex corrections are not taken into account in our calculation. 

\subsection{Role of band broadening}

We discuss the dependence on the phenomenological relaxation rate $\Gamma$, which is captured by the spectral weighting factors \eqref{eqn:wintra}-\eqref{eqn:winterA}. We distinguish three different cases: the clean limit for a metal, charge neutrality points in (higher-order) Dirac semimetals, and flat bands. We note that the scaling behavior for large $\Gamma$ and the (topological) properties of the Berry curvature contribution $\sigma^{\alf\beta,a}_\text{inter}$ were already discussed in Ref.~\cite{Mitscherling2020} and will not be considered further here.

\paragraph{Clean limit for a metal}
In the following, we generalize the results for two-band systems given in Refs.~\cite{Mitscherling2020, Mitscherling2018}. We assume a $d-1$-dimensional Fermi surface. If $\Gamma$ is so small that for given directions $\alf$, $\beta$, and bands $n,m$ the quantities $v^{\alf,n}_\bp v^{\beta,n}_\bp$, $g^{\alf\beta,nm}_\bp$, and $\Omega^{\alf\beta,nm}_\bp$ are almost constant in a momentum range in which the variation of $E^n_\bp$ and $E^m_\bp$ is order $\Gamma$, we can approximate the spectral weighting factors in Eqs.~\eqref{eqn:wintra}-\eqref{eqn:winterA} by Dirac delta functions with a particular leading-order dependence on $\Gamma$: $w^n_{\bp,\text{intra}}\sim 1/\Gamma$, $w^{nm,s}_{\bp,\text{inter}}\sim \Gamma$, and $w^{nm,a}_{\bp,\text{inter}}\sim 1$.
After performing the frequency integration and one of the band traces, we see that the intraband, the quantum metric and the Berry curvature contributions further decompose into individual band contributions $\sigma^{\alf\beta}_{\text{intra},n}\sim v^{n,\alf}_\bp v^{n,\beta}_\bp$, $\sigma^{\alf\beta,s}_{\text{inter},n}\sim g^{\alf\beta,n}_\bp$, and $\sigma^{\alf\beta,a}_{\text{inter},n}\sim \Omega^{\alf\beta,n}_\bp$, respectively, involving the respective quasiparticle velocities, quantum metric, and Berry curvature. Thus we recover the well-known semiclassical results for the intraband contribution \cite{Ashcroft1976} as well as the dissipationless intrinsic anomalous Hall conductivity due to the Berry curvature \cite{Nagaosa2010, Xiao2010}. The quantum metric contribution of the longitudinal conductivity is suppressed by $\Gamma^2/(E^n_\bp-E^m_\bp)^2$ compared to the intraband contribution. It provides significant contribution only at small direct band gaps, for instance, at the onset of order at quantum critical points \cite{Mitscherling2018, Bonetti2020}.

\paragraph{Charge neutrality points in Dirac semimetals.} Let us first give a few examples for which the intraband and the quantum metric contributions are of the same order in $\Gamma$. For two- and three-dimensional Dirac cones we find the conductivity independent of $\Gamma$. This also holds for a nodal dispersion with quadratic band touching in two dimensions, whereas we find an unconventional scaling of $1/\sqrt{\Gamma}$ in $x$ direction and $\sqrt{\Gamma}$ in $y$ direction for a mixed linear and quadratic band touching. In all these cases, the size of the quantum metric contribution is of the same order as the intraband contribution. For instance, both $\sigma^{xx}_\text{intra}$ and $\sigma^{xx,s}_\text{inter}$ contribute equally to the longitudinal conductivity of a two-dimensional Dirac cone. Nevertheless, the ratio between the quantum metric and the intraband contribution is nonuniversal as explicitly shown for the model of quadratic band touching. A table which summarizes these results can be found in  Appendix~\ref{sec:Semimetals}.

\paragraph{Flat-band geometry.}

We study the intraband and the quantum metric contribution for a flat-band system. For convenience, we relabel the bands such that the zeroth band is the momentum-independent flat band, $E^0_\bp=E_\text{flat}$. 
We fix the chemical potential at the flat-band energy $\mu=E_\text{flat}$ and assume that the flat band is isolated such that $\Gamma\ll|E^n_\bp-E_\text{flat}|$ for all bands $n\neq 0$ and momenta $\bp$. Note that the following arguments also hold for bands $E^0_\bp$ that are flat on the scale of $\Gamma$, i.e., $|\mu-E^0_\bp|\ll\Gamma$ for all momenta $\bp$. For improved clarity, we will present all results for zero temperature. The spectral weighting factors involving the flat band are 
%
%\begin{linenomath*}
\begin{align}
    &w^0_{\bp,\text{intra}}(0)\approx \frac{1}{\pi\Gamma^2}\label{eqn:WintraFlat}\,,\\
    &w^{0n,s}_{\bp,\text{inter}}(0)=w^{n0,s}_{\bp,\text{inter}}(0)\approx \frac{1}{\pi}\label{eqn:WinterSFlat}
\end{align}
%\end{linenomath*}
%
for $n\neq0$. The higher-order terms and the other contributions with $n,m\neq0$ are of order  $\Gamma^2/(E^n_\bp-E_\text{flat})^2$ and, thus, suppressed. Using the formula in Eq.~\eqref{eqn:SinterS}, we see that the quantum metric contribution is given by the integral of the quantum metric over the Brillouin zone, 
%
%\begin{linenomath*}
\begin{align}
    \label{eqn:SinterSflat}
    \sigma^{\alf\beta,s}_\text{inter}=\frac{2}{\pi}\frac{e^2}{\hbar}\hspace{-1mm}\int\hspace{-1mm}\frac{d^d\bp}{(2\pi)^d}\, \,g^{\alf\beta,0}_\bp \, .
\end{align}
%\end{linenomath*}
%
Rewriting the quantum metric as $g^{\alf\beta,n}_\bp=\langle n_\bp|\hat r_\alf\hat r_\beta|n_\bp\rangle-\langle n_\bp|\hat r_\alf|n_\bp\rangle\langle n_\bp|\hat r_\beta|n_\bp\rangle$ with position operator $\hat r_\alf=i\partial_\alf$ we see that $\sigma^{\alf\beta,s}_\text{inter}$ for a chemical potential inside the flat band is given by the mean spread of the Bloch wave functions of the flat band \cite{Marzari1997, Matsuura2010, Marzari2012, Wang2020a}.

We now focus on a two-dimensional system, which we assume to lie in the $x$-$y$ plane. The trace of the quantum metric is bounded by the absolute value of the Berry curvature \cite{Roy2014a, Peotta2015, Ozawa2021}, that is, $g^{xx,n}_\bp+g^{yy,n}_\bp\geq|\Omega^{xy,n}_\bp|$, since the quantum geometric tensor $\mathcal{T}^{\alf\beta}_\bp$ is positive semi-definite. Thus the sum of the quantum metric contribution in $x$ and $y$ direction is bounded from below, that is, 
%
%\begin{linenomath*}
\begin{align}
    \sigma^{xx,s}_\text{inter}+\sigma^{yy,s}_\text{inter}
    \geq\frac{2}{\pi}\frac{e^2}{\hbar}\hspace{-1mm}\int\hspace{-1mm}\frac{d^d\bp}{(2\pi)^d}\, \,|\Omega^{xy,0}_\bp|\geq\frac{2}{\pi}\frac{e^2}{2\pi\hbar}|C_0| \, , \label{eqn:bound}
\end{align}
%\end{linenomath*}
%
where the Chern number is defined by  $C_n=2\pi\hspace{-1mm}\int\hspace{-1mm}\frac{d^2\bp}{(2\pi)^2}\Omega^{xy,n}_\bp\in \mathds{Z}$. We see that the quantum metric contribution and thus the longitudinal conductivity is bounded for systems with nonzero Berry curvature. 

The intraband contribution \eqref{eqn:Sintra} involves the quasiparticle velocities, which are identically zero in a perfect flat band. Thus the conductivity is entirely given by the quantum metric contribution. In almost flat bands, both intra- and interband processes contribute in general. Due to the different scaling in $\Gamma$ [see Eqs.~\eqref{eqn:WintraFlat} and \eqref{eqn:WinterSFlat}] the intraband contribution can still dominate for sufficiently small $\Gamma$. We can give an estimate on the crossover scale by setting $\sigma^{xx}_\text{intra}+\sigma^{yy}_\text{intra}=\sigma^{xx,s}_\text{inter}+\sigma^{yy,s}_\text{inter}$, from which we conclude that the interband contribution is dominant for values of the band broadening above a threshold $\Gamma_c$, with $\Gamma_c\lesssim \big(\pi/|C_0|\big[( v^{x,0}_\text{max})^2+(v^{y,0}_\text{max})^2\big]\big)^{1/2}$ 
%
%\begin{linenomath*}
%\begin{align}
%    \Gamma_c\lesssim \sqrt{\frac{\pi}{|C_0|}\Big[( v^{x,0}_\text{max})^2+(v^{y,0}_\text{max})^2\Big]} 
%\end{align}
%\end{linenomath*}
%
by using \eqref{eqn:bound} and approximating the quasiparticle velocities by their upper bounds $v^{\alf,0}_\text{max}$. However, note that a large estimate for $\Gamma_c$ does not imply a negligible interband contribution for smaller $\Gamma$. We find that the estimate of $\Gamma_c$ becomes less indicative of the crossover scale with increasing bandwidth of the (nearly) flat band. In particular, in our examples $\Gamma_c$ can be used to understand the result of the flat-band toy model but is less helpful in the description of rhombohedral trilayer graphene.

\subsection{Flat band model}

\begin{figure}
    \centering
    \includegraphics[width=0.235\textwidth]{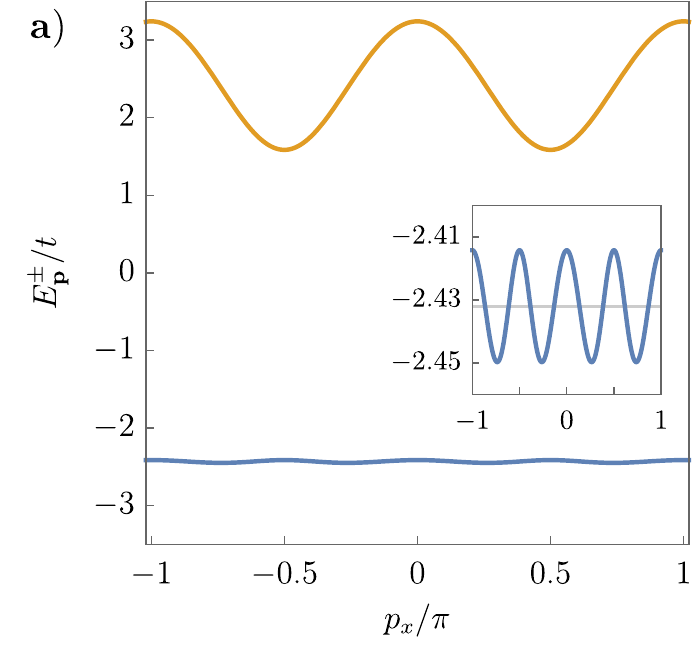}
    \includegraphics[width=0.235\textwidth]{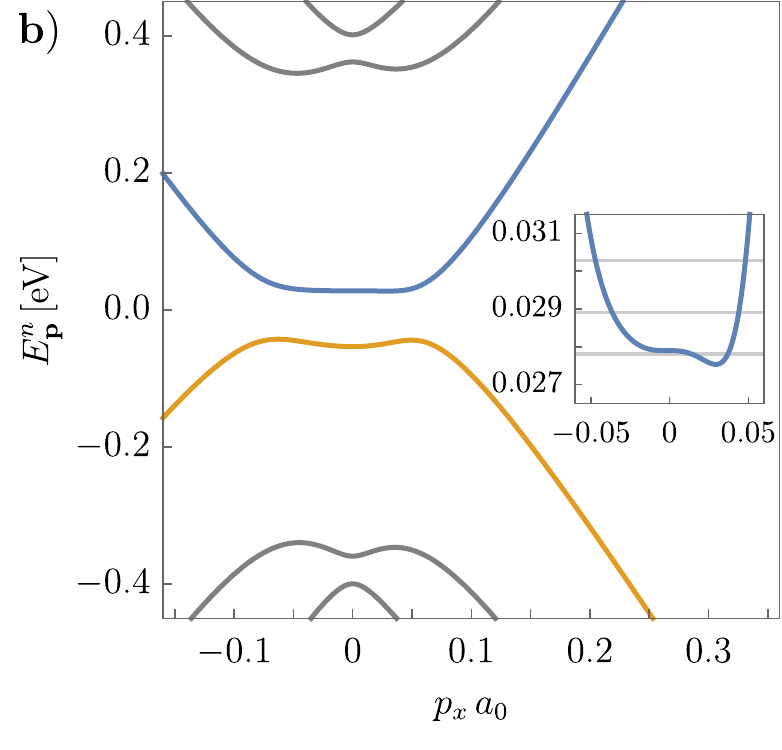}
    %\internallinenumbers
    \caption{The dispersions of the flat-band model (a) and of the model for rhombohedral trilayer graphene (b). The dispersion of the considered flat band (blue) and the chemical potentials are shown in the inset. The band width is much smaller than the gap to the closest band (orange).}
    \label{fig:dispersion}
\end{figure}

\begin{figure}
    \centering
    \includegraphics[width=0.4\textwidth]{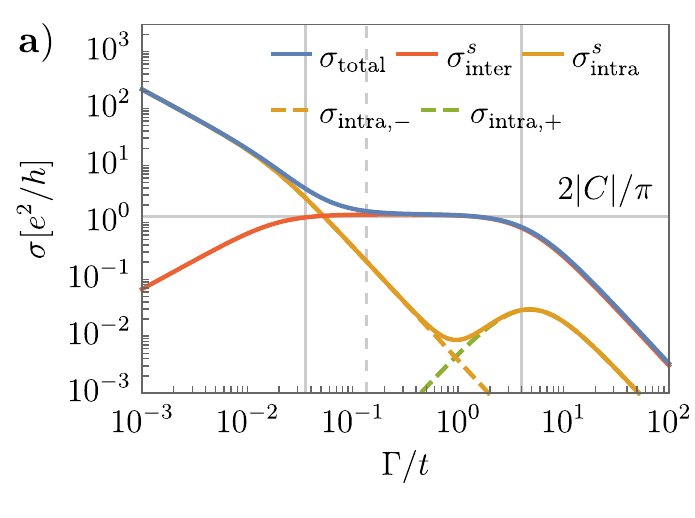}
    \includegraphics[width=0.4\textwidth]{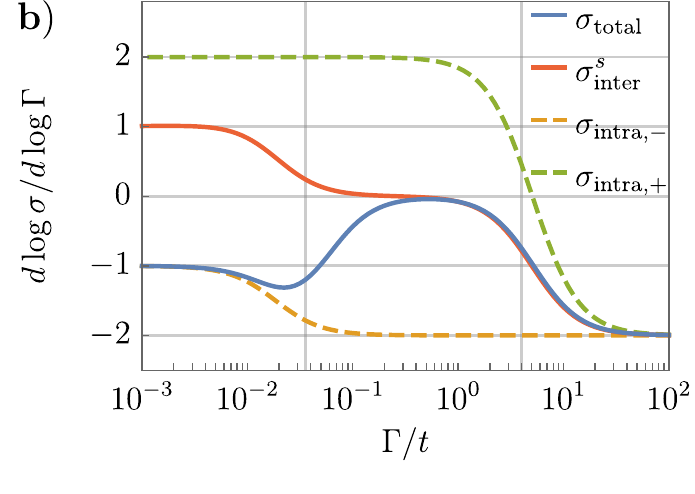}
    %\includegraphics[width=0.3\textwidth]{figures/fig3c.pdf}
    %\internallinenumbers
    \caption{The longitudinal conductivity for the flat band model with chemical potential fixed to the center of the lower and (almost) flat band, $\mu=(\mu^\text{max}_-+\mu^\text{min}_-)/2=-2.432\,t$. (a) The total conductivity $\sigma^{xx}_\text{total}+\sigma^{yy}_\text{total}$ (blue) with its intraband (orange) and geometric interband contribution (red) as a function of $\Gamma/t$. The conductivity is dominated by the geometric interband contribution beyond $\Gamma_c/t\approx 0.134$ (dashed vertical line), which is bounded from below (horizontal line) due to the finite Chern number $|C|=2$, in an intermediate regime approximately given by the band width of the flat band $W_-/t=0.036$ and the band gap $\Delta/t=4$ (solid vertical lines). At small $\Gamma$ the conductivity is dominated by the intraband contribution of the (almost) flat band (dashed orange). As expected, the bound ceases to be effective for very large $\Gamma$. (b) The scaling behavior of the different contributions. At low and large $\Gamma$, the total conductivity and its contributions scale as expected \cite{Mitscherling2020}. In the intermediate regime the conductivity is independent of $\Gamma$.}
    \label{fig:flatband}
\end{figure}

We apply our theory to a two-band model hosting a topological nontrivial flat lower and a dispersive upper band which was introduced in Ref.~\cite{Hofmann2020}. We show the band structure in Fig.~\ref{fig:dispersion}. The complete Hamiltonian, the explicit form of the two bands $E^\pm_\bp$ as well as the quantum metrics $g^{xx,\pm}_\bp$ and $g^{yy,\pm}_\bp$ are provided in Appendix~\ref{sec:flatband}. We use $t$ as our unit of energy and set the lattice constant $a=1$. We fix the chemical potential to the center of the lower band and analyze the longitudinal conductivity and its different contributions in the following. In Fig.~\ref{fig:flatband}(a), we show the total longitudinal conductivity $\sigma_\text{total}=\sigma^{xx}+\sigma^{yy}$ (blue) calculated via \eqref{eqn:sigma} at zero temperature as a function of $\Gamma/t$. We have $\sigma^{xx}=\sigma^{yy}$ as expected by the $C_4$ symmetry of the model. We clearly identify three different regimes with the crossover scales given by the band width of the flat band $W_-/t=0.036$ (left vertical gray line) and the band gap $\Delta/t=4$ (right vertical gray line). For small $\Gamma\lesssim W_-$ the conductivity is dominated by the intraband contribution \eqref{eqn:Sintra} (orange) due to the small but finite quasiparticle velocities with maximal values $v^{x,-}_\text{max}=v^{y,-}_\text{max}=0.076\,t$ of the flat band (dashed orange) . For an intermediate relaxation rate, $W_-\lesssim \Gamma\lesssim \Delta$, the quantum metric contribution \eqref{eqn:SinterS} (red) is independent of $\Gamma$ and bounded from below by the finite Chern number $C_-=2$ of the flat band according to \eqref{eqn:SinterSflat} and \eqref{eqn:bound}, respectively. Above the threshold $\Gamma_c/t\approx 0.134$ (dashed vertical gray line) the total conductivity is dominated by the quantum metric contribution. For large $\Gamma\gtrsim \Delta$ (right vertical gray line), the interband contribution is no longer given by Eq.~\eqref{eqn:SinterSflat}, so that the lower bound \eqref{eqn:bound} no longer applies. If $\Gamma$ exceeds the band width of all bands, $\Gamma/t\gg(E^+_\text{max}-E^-_\text{min})/t\approx5.7>\Delta/t $, both the intraband and the interband contribution are strongly suppressed like $\Gamma^{-2}$ \cite{Mitscherling2020}. 

In this simple model, the integral of Eq.~\eqref{eqn:SinterSflat} over the quantum metric can be calculated explicitly, yielding
\begin{align}\label{eq:explicitg}
    &\int \frac{d^2\bp}{(2\pi)^2}\big(g^{xx,-}_\bp+g^{yy,-}_\bp\big)=\frac{K(-1/8)}{\sqrt{2}\pi}>\frac{|C_-|}{2\pi}\, ,
\end{align}
where $K(m)=\int_0^{\pi/2}\big(1-m\sin^2\theta\big)^{-1/2}d\theta$ is the complete elliptic integral of the first kind. 
The numerical value of Eq.~\eqref{eq:explicitg} is $0.34$, less than 10\% above the threshold $1/\pi\approx 0.32$. We see that in a nearly perfect flat band, the bound holds tight.

In Fig.~\ref{fig:flatband}(b), we plot the logarithmic derivative of the different conductivity contributions as a function of $\Gamma/t$. For $\Gamma\ll W_-$ we have $\sigma_\text{intra,-}\sim 1/\Gamma$ and $\sigma^s_\text{inter}\sim \Gamma$ as expected for the clean limit behavior of a metal as discussed above. The intraband contribution of the upper band scales as $\sigma_{\text{intra},+}\sim \Gamma^2$ \cite{Mitscherling2020}. For $\Gamma\gg \Delta$, we have $\sigma_{\text{intra},\pm}\sim\sigma^s_\text{inter}\sim1/\Gamma^2$ \cite{Mitscherling2020}. In the intermediate regime $W_-\ll \Gamma\ll \Delta$, we have $\sigma_{\text{intra},-}\sim 1/\Gamma^2$ and $\sigma^s_\text{inter}\sim 1$ as expected by \eqref{eqn:WintraFlat} and \eqref{eqn:WinterSFlat}, respectively.

\subsection{Rhombohedral trilayer graphene}
\label{sec:RTG}

We apply our theory to a realistic six-band low-energy model of two-dimensional rhombohedral trilayer graphene in the regime of low carrier density where the fourfold flavor-symmetry in valley and spin is broken and only one flavor is occupied~\cite{Zhou2021}. The full Hamiltonian and the parameters, which we adapted from Ref.~\cite{Zhou2021}, can be found in Appendix~\ref{sec:RTGappendix}. The material is highly tunable via an electrical displacement field, which introduces an energy gap $\Delta_1$. The band structure is shown in Fig.~\ref{fig:dispersion}. Bands 3 (orange) and 4 (blue) are almost flat within a momentum range of $|\bp|/a_0\lesssim 0.05$, where $a_0$ is the lattice constant. In the following, we probe the longitudinal conductivity at the lower edge $\mu_{4,\text{min}}$ of the nearly flat conduction band 4. The integration was performed on a square region up to a momentum cutoff $\pi a_0\gg 0.05a_0$. 

\begin{figure}
    \centering
    \includegraphics[width=0.4\textwidth]{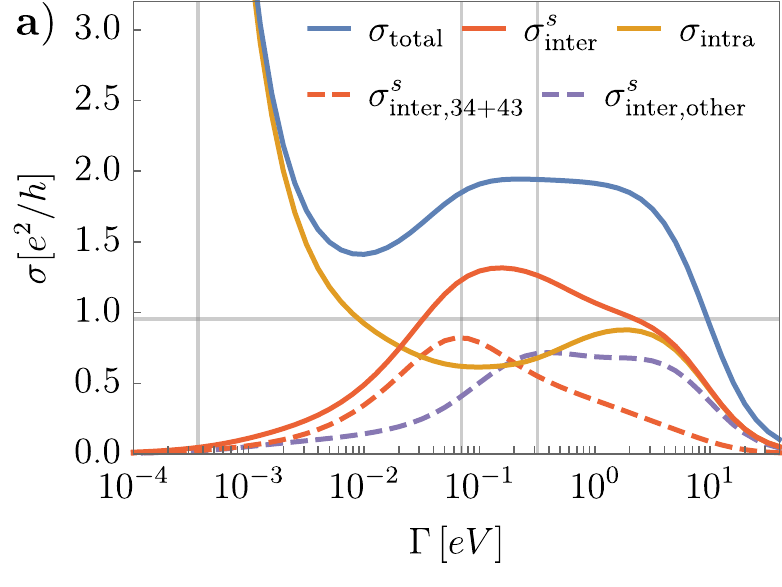}\\[2mm]
    \includegraphics[width=0.4\textwidth]{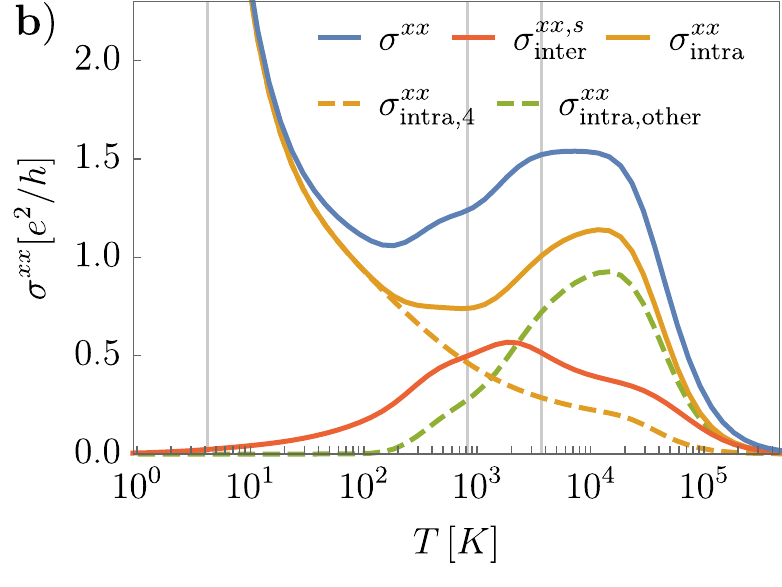}
    %\internallinenumbers
    \caption{The contributions to the total longitudinal conductivity of rhombohedral trilayer graphene. (a) The total conductivity (blue) and its intraband (orange) and geometric interband (red) contributions as a function of $\Gamma$ for $\mu=1.05\,\mu_{4,\text{min}}$ and $\Delta_1=40\,\text{meV}$ at $T=0$. The main contribution to $\sigma^s_\text{intra}$ at intermediate $\Gamma$ are due to interband couplings of bands $3$ and $4$ (red dashed) but all other 28 couplings contribute (purple dashed) significantly.  
    (b) The total (blue) contributions and its intraband (orange) and geometric interband (red) contributions as a function of temperature $T$ for the same parameters with $\Gamma=k_B T$. At low $T$, the intraband contribution is dominated by band $4$ (orange dashed). At higher $T$, the other bands (purple dashed) contribute significantly. The geometric contribution causes a conductivity minimum within the temperature range $T\approx 50...400\,K$.}
    \label{fig:trilayer}
\end{figure}

In Fig.~\ref{fig:trilayer}(a), we show the total conductivity $\sigma_\text{total}=\sigma^{xx}+\sigma^{yy}$ (blue) and its quantum metric (red) and intraband (orange) contribution as a function of $\Gamma$ at zero temperature. We fix $\Delta_1=40\,\text{meV}$ and $\mu=\mu_{4,\text{min}}+1.4\text{meV}$. Although the band structure is more complicated than for the flat-band model, we find similar characteristic regimes. For intermediate relaxation rates between $\Gamma\approx 0.03...3\,\text{eV}$ there is a large geometric interband contribution exceeding the lower bound \eqref{eqn:bound} due to the Berry charges totaling $C_4=3/2$ near the band bottom of the fourth band. For small values of $\Gamma$, the main part of the quantum metric contribution is due to the interband coupling of the close bands $3$ and $4$, i.e., $\sigma^s_\text{inter}\approx \sigma^s_{\text{inter},34}+\sigma^s_{\text{inter},43}$. For larger $\Gamma$ the quantum metric contribution of all other $28$ interband couplings $\sigma^s_{\text{inter},\text{other}}$ are required to exceed the lower bound. This might become important when using projections on a single band. Note that in contrast to the flat-band model discussed before, the intermediate regime cannot be easily connected to the characteristic energy scales of the system (vertical gray lines for the band width of the fourth band $W_4$ and the gaps to the third and fifth band $\Delta_{34}$ and $\Delta_{45}$, respectively). The main reason is the finite momentum range in which the fourth band remains flat: The relevant spectral weighting factors $w^{4n,s}_{\bp,\text{inter}}(0)$ and $w^{n4,s}_{\bp,\text{inter}}(0)$ with band index $n\neq 4$ of the other five bands fulfill  Eq.~\eqref{eqn:WinterSFlat} only for momenta $|\bp|/a_0\lesssim 0.05$ and are strongly suppressed otherwise. Thus the range of the momentum integration in \eqref{eqn:SinterSflat} is restricted and does not capture the full quantum metric, which is maximal for $0.03\lesssim (g^{xx,4}_\bp+g^{yy,4}_\bp)/a_0\lesssim 0.12$. The maximal value of $\sigma^s_\text{inter}$ at $\Gamma\approx 0.2\,\text{eV}$ is due to a subtle interplay between the spectral weighting factors and the quantum metric in \eqref{eqn:SinterS}. The upper scale is given by the momentum cutoff. At very low $\Gamma\sim W_4=0.4\,\text{meV}$ of the order of the band width of the flat band $W_4$, the conductivity is dominated by the intraband contribution as expected due the existence of a Fermi surface. Due to the upturn of the flat band at $|\bp|/a_0\approx 0.05$ and the corresponding finite quasiparticle velocities, the interband contribution is nonzero over the full range of $\Gamma$, in particular, when $\Gamma$ exceeds the gaps to bands $3$ and $5$ with $\Delta_{34}=70\,\text{meV}$ and $\Delta_{45}=317\,\text{meV}$. The upper scale is again given by the momentum cutoff. 

The temperature dependence of the conductivity in RTG beyond $100\mathrm{K}$ has not yet been investigated in detail, but the presence of flat bands and the emergence of superconducting phases at low temperatures suggests that the material offers a similar phenomenology to TBG. The latter has been shown to exhibit a very large linear-$T$ resistivity, possibly making it a Planckian metal~\cite{Cao2020} in which the relaxation rate is close to a putative upper bound $\mathcal{O}(k_B T)$.
Therefore, for the discussion of the temperature dependence and possible experimental signatures in RTG, we feel it is justified to assume the "best case scenario" for our purposes, where $\Gamma\sim k_B T$ with a coefficient close to 1.
In Fig.~\ref{fig:trilayer}(b), we show the conductivity $\sigma^{xx}$ (blue) and its quantum metric (red) and intraband (orange) contributions as a function of temperature. We used the same parameters for $\Delta_1$ and $\mu$ as in the upper figure and assumed a relaxation rate of $\Gamma=k_B T$. We obtain $\sigma^{xx}=\sigma^{yy}$ within our numerical accuracy.  According to Eqs.~\eqref{eqn:Sintra} and \eqref{eqn:SinterS} a finite temperature enters not only through the band broadening, but also via the Fermi-Dirac function in the spectral weights. Nevertheless, the characteristic behavior of the different contributions are qualitatively similar to the zero temperature case. In the low-temperature regime, $T\sim W_4/k_B\approx 5\,K$, the conductivity is mainly given by the intraband contribution of the fourth band (orange and dashed), which can be understood by the scaling argument $\sigma^{xx}_\text{intra}\sim 1/\Gamma\sim 1/T$. At high temperatures beyond $T\sim \Delta_{34}/k_B\approx800\, K$, the other bands contribute significantly (green and dashed). In the intermediate range, the quantum metric contribution becomes crucial. We see that the temperature broadening by the Fermi function enhances the intraband contribution, whereas the quantum metric contribution is little affected. 

\begin{figure}
    \centering
    \includegraphics[width=0.4\textwidth]{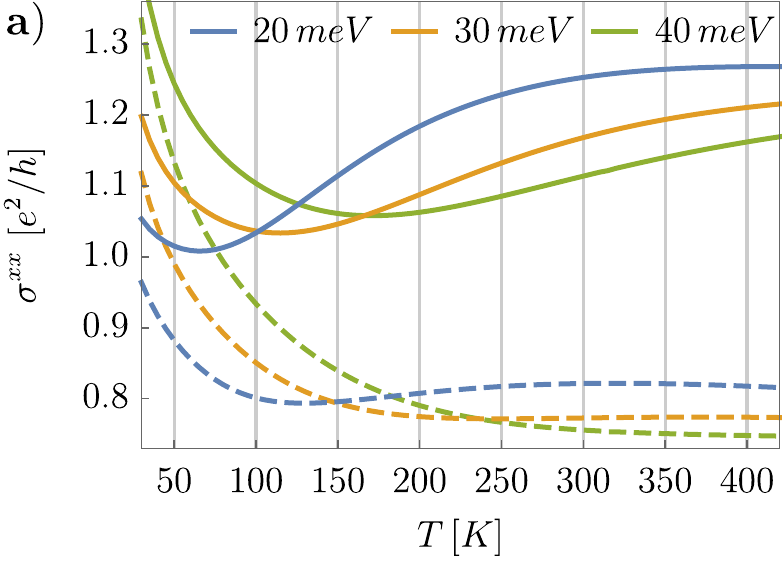}\\[2mm]
    \includegraphics[width=0.4\textwidth]{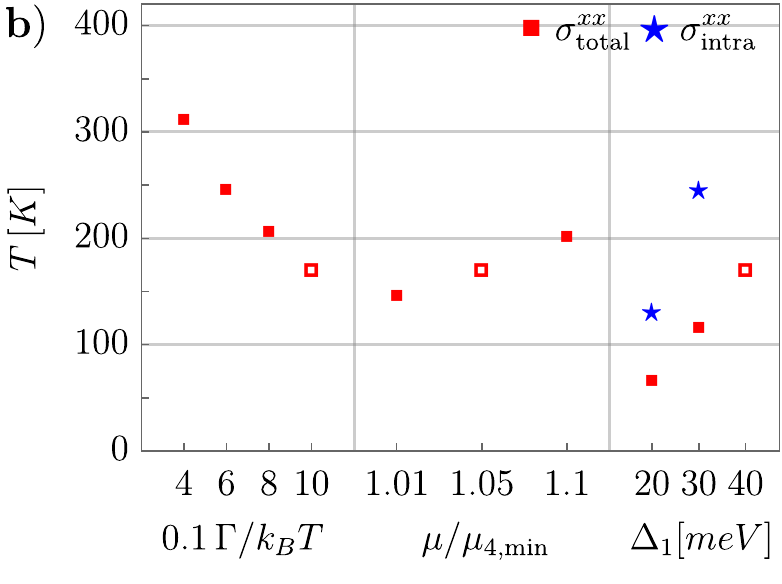}
    %\internallinenumbers
    \caption{The quantum-metric induced conductivity minimum in trilayer graphene. 
    (a) The total (solid) and the intraband (dashed) conductivity as a function of temperature for different $\Delta_1$ at fixed $\mu=1.05\,\mu_{4,\text{min}}$ and $\Gamma=k_B T$. 
    (b) The temperature of the minimal total (red filled and empty box) and intraband (blue star) conductivity for different parameters compared to $\mu=1.05\,\mu_{4,\text{min}}$,  $\Gamma=k_B T$, and $\Delta_1=40\,\text{meV}$ (empty box).} 
    \label{fig:minimum}
\end{figure}

The conductivity $\sigma^{xx}$ exhibits a minimum at approximately $T_\text{min}\approx 170\,K$. The decomposition into its contributions shown in Fig.~\ref{fig:trilayer} reveals the origin of this minimum. It is caused by the decrease of the intraband contribution of the almost flat fourth band and the increase of the quantum metric contribution. It is important to notice that the thermal activation of the other bands leading to further intraband contributions is not sufficient to a cause a minimum at such a low energy scale $k_B T_\text{min}\sim 15\,\text{meV}$, which is much smaller than the smallest gap $\Delta_{34}=70\,\text{meV}$. The intraband contributions of the other bands lead to a second kink at much higher temperature. Thus, we have identified a minimum of the longitudinal conductivity due to virtual processes, i.e., effects of the quantum metric in rhombohedral trilayer graphene. 

In Fig.~\ref{fig:minimum}, we discuss the stability and trends of the minimum for different parameter values. In Fig.~\ref{fig:minimum}(a), we show $\sigma^{xx}$ (solid) and the corresponding intraband contribution (dashed) for different $\Delta_1$ at fixed $\Gamma=k_B T$ and $\mu/\mu_{4,\text{min}}=1.05$, that is $\mu-\mu_{4,\text{min}}=1.4,\,0.9,\,0.4\,\text{meV}$ for $\Delta_1=40,\,30,\,20\,\text{meV}$. We see a pronounced minimum for all three parameter sets. The comparison to the intraband contribution clearly shows that this minimum is caused by virtual processes captured by the quantum metric and not by thermal activation of intraband processes. Only a small minimum at signficiant higher temperature can be found for $\Delta_1=20\,\text{meV}$ (blue) and $\Delta_1=30\,\text{meV}$ (orange). We summarize the temperature of the conductivity minimum in Fig.~\ref{fig:minimum}~b) for these and further parameters of $\Gamma/k_B T$, $\mu/\mu_{4,\text{min}}$, and $\Delta_1$. The minimum found in Fig.~\ref{fig:trilayer}~b) is taken as a reference (empty red box). We see that decreasing the temperature dependence of $\Gamma$ shifts the minimum to higher temperature. We understand this shift by both an increase of the intraband and a decrease of the quantum metric contribution by the following argument: The conductivity minimum at temperature $T_\text{min}$ satisfies $d\sigma^{xx}/dT=0$. Assuming the scaling $\sigma^{xx}_\text{intra}\sim 1/ak_BT$ and $\sigma^{xx,s}_\text{inter}\sim ak_BT$ leads to $T_\text{min}\sim 1/a$ with $a=0.4,\,0.6,\,0.8,$ and $1.0$. Note that the assumed scaling is only expected for small temperature. An increase of the chemical potential $\mu/\mu_{4,\text{min}}=1.01,\,1.05,$ and $1.1$, that is $\mu-\mu_{4,\text{min}}=0.3,\,1.4,$ and $2.8\,\text{meV}$, respectively, increases the Fermi velocity on the Fermi surface [cf. Fig.~\ref{fig:dispersion}]. This increases the intraband contribution \eqref{eqn:Sintra}, whereas the quantum metric contribution remains unchanged. Thus the minimum shifts to higher temperatures. Modifying $\Delta_1$ decreases the gap between bands $3$ and $4$, which increases the quantum metric. This and a decrease of the intraband contribution shifts the minimum to lower temperature. Whereas the full conductivity $\sigma^{xx}$ exhibits a conductivity minimum for all considered parameters, the intraband contribution $\sigma^{xx}_\text{intra}$ does not show any minimum in the reasonable temperature range for most of the parameters.

We end with a discussion of the limitations of our approach.
Since the formalism makes exclusive use of the band broadening parameter $\Gamma$, a legitimate concern is whether vertex corrections and extrinsic scattering due to impurities can qualitatively affect the results presented here. 
While both these effects can indeed become problematic in semimetals which necessarily have a small density of states and a large Fermi velocity~\cite{Sinitsyn2007,Nagaosa2010}, for a flat band dispersion it is the opposite, as it has a large density of states and a vanishing quasiparticle velocity. It therefore behaves like ordinary metals with a large Fermi surface, where the quantum lifetime and the transport lifetime are identical and impurities effects are weak.

We also limited the analysis to the non-interacting case. Upon introducing interactions, additional interband processes become possible which are mediated by these interactions. Since this introduces additional virtual transitions, we expect it to increase both the effective $\Gamma$ and the effective $g$. As a result, it could be that the bound on resistivity is only weakly fulfilled. Since the bound is already only weakly fulfilled due to the finite residual Fermi velocity in RTG, it does not affect our main conclusions.
A more serious complication is the possible presence of ordered states near charge neutrality. Experimentally, a correlated insulating state seems to emerge for carrier densities below $n_e=0.05\times 10^{-12}\mathrm{cm^{-2}}$, corresponding to a chemical potential of $1$-$2\,\mathrm{meV}$ above the band termination, which is comparable to the densities considered here. It is presently unclear whether such a state will also exist in samples of lower mobility.

\section{Conclusions}
We systematically explored the role of the quantum metric for the electrical conductivity for various semimetallic and metallic model systems, finding that it contributes at leading order both if the density of states vanishes at semimetallic band crossing points, but also when the density of states diverges in systems with quenched bandwidth and vanishing Fermi velocity. 
We reported an upper bound on the resistivity for flat-band materials with nonzero Berry curvature and elucidated the range of validity and why this bound is limited to an intermediate range of relaxation rates.
As a platform to explore the resistivity bound, we suggested to employ rhombohedral trilayer graphene, and gave a detailed account of the various crossover scales which are expected in the system. 
After completion of this work, the simultaneously finished Ref.~\cite{HerzogArbeitman2021} was brought to our attention, who identified lower bounds to the quantum metric in flat-band systems whose Wannier centers are obstructed from the atoms and which do not necessarily have a finite Berry curvature. This result suggests that the bound on the resistivity due to the quantum metric applies to a much broader class of materials as long as our other assumptions are fulfilled.

It is of timely relevance to explore how interactions will change the phenomenology reported here, in particular with respect to the quantum metric, a band structure parameter like the Berry curvature which is probably quite robust against interaction effects.

%%%%%%%%%%%%%%%%%%%%%%%%%%%%%%%%%%%%%%%%%%%%%%%%%%%%%%%%%%%%%%%%%%%%%%%%%%%%%%%%%%%%
\begin{acknowledgments}
We are very grateful to Erez Berg, Laura Classen,
Johannes Hofmann, 
and Elio K\"onig
for valuable discussions.
\end{acknowledgments}

%%%%%%%%%%%%%%%%%%%%%%%%%%%%%%%%%%%%%%%%%%%%%%%%%%%%%%%%%%%%%%%%%%%%%%%%%%%%%%%%%%%%%%%%

\appendix
%\section*{Appendix}

\section{Kubo formalism for fast relaxation} \label{sec:KuboFormalism}

We calculate the dc conductivity given by
%
%\begin{linenomath*}
\begin{align}
    \sigma^{\alf\beta}=-\lim_{\omega\rightarrow 0}\frac{1}{i\omega}\Pi^{\alf\beta}(\omega) \, ,
\end{align}
%\end{linenomath*}
%
where $\alf$ and $\beta$ denote the spatial directions of the current and the electric field, respectively, in $d$ dimensions for the Hamiltonian given in Eq.~\eqref{eqn:H}. The following derivation generalizes the approach developed by one of us from two to $N$ bands \cite{Mitscherling2020}. The polarization tensor $\Pi^{\alf\beta}(\omega)$ is obtained by the Kubo formula
%
%\begin{linenomath*}
\begin{align}
    \label{eqn:Pi}
    \Pi^{\alf\beta}_{iq_0}=\frac{e^2}{\hbar}\frac{k_B T}{V}\hspace{-1mm}\sum_{ip_0,\bp}\hspace{-1mm}\tr\big[\sG^{}_{ip_0+iq_0,\bp}\lam^\beta_\bp\sG^{}_{ip_0,\bp}\lam^\alf_\bp-(iq_0=0)\big]
\end{align}
%\end{linenomath*}
%
after analytic continuation of the bosonic Matsubara frequency $iq_0\rightarrow \omega+i0^+$ to real frequency $\omega$. The polarization tensor \eqref{eqn:Pi} involves the generalized velocities, that is, the momentum derivatives of the Bloch matrix in Eq.~\eqref{eqn:H}, $\lam^\alf_\bp=\partial_\alf\lam_\bp$, and the Green's function matrix 
%
%\begin{linenomath*}
\begin{align}
    \sG_{ip_0,\bp}=\big[ip_0+\mu-\lam_\bp+i\Gamma \text{sign}(p_0)\big]^{-1}
\end{align}
%\end{linenomath*}
%
with fermionic Matsubara frequency $ip_0$. We denote the trace over the $N$ bands as $\tr$. $T$ is the temperature and $V$ the volume of the Brillouin zone. The second term in Eq.~\eqref{eqn:Pi} is the diamagnetic term, which is equal to the first term with $iq_0$ set to zero.

The key in the used approach is to consider a phenomenological relaxation rate $\Gamma>0$, which is not restricted in size in the following derivation \cite{Mitscherling2020}. For simplicity, $\Gamma$ is assumed to be frequency- and momentum-independent as well as equal for all bands. 

We diagonalize the Bloch Hamiltonian $\lam_\bp$ by its $N$ eigenvectors $|n_\bp\rangle$ with corresponding eigenvalues $E^n_\bp$. The elements of the Green's function and the generalized velocities in the eigenbasis then read
%
%\begin{linenomath*}
\begin{align}
    \label{eqn:Gn}
    &\cG^n_{ip_0,\bp}=\big[ip_0+\mu-E^n_\bp+i\Gamma\text{sign}(p_0)\big]^{-1} \, ,\\[1mm]
    \label{eqn:dLamNM}
    &\big(\tilde \lam^\alf_\bp\big)_{nm}=\delta_{nm}v^{\alf,n}_\bp+i(E^n_\bp-E^m_\bp)\cA^{\alf,nm}_\bp \, ,
\end{align}
%\end{linenomath*}
%
where we denote the quasiparticle velocities as $v^{\alf,n}_\bp=\partial_\alf E^n_\bp$ and the Berry connection as $\cA^{\alf, nm}_\bp=i\langle n_\bp|\partial_\alf m_\bp\rangle$. $\delta_{nm}$ is the Kronecker delta. Note that the quasiparticle Green's function \eqref{eqn:Gn} is diagonal, whereas the generalized velocity \eqref{eqn:dLamNM} has both diagonal and off-diagonal contributions leading to both intraband and interband contributions to the conductivity tensor $\sigma^{\alf\beta}$. Furthermore, we uniquely decompose $\sigma^{\alf\beta}$ into its symmetric and antisymmetric parts under the exchange of $\alf\leftrightarrow\beta$. After performing the Matsubara summation, analytic continuation, and the thermodynamic limit $1/V\sum_\bp\rightarrow \int\hspace{-1mm}d^d\bp/(2\pi)^d$, we obtain Eq.~\eqref{eqn:sigma} with the three contributions \eqref{eqn:Sintra}-\eqref{eqn:SinterA}.

Note that the two quantities $g^{\alf\beta,nm}_\bp$ and $\Omega^{\alf\beta,nm}_\bp$ defined in \eqref{eqn:gnm} and \eqref{eqn:Onm} are invariant under the $U(1)$ gauge freedom $|n_\bp\rangle\rightarrow e^{i \phi_\bp}|n_\bp\rangle$ for $n\neq m$ due to the particular combination of the Berry connections. The summation $\sum_{n\neq m}\cA^{\alf,nm}_\bp\cA^{\beta,mn}_\bp$ yields the quantum geometric tensor $\mathcal{T}^{\alf\beta,n}_\bp$, which decomposes into the symmetric real part $g^{\alf\beta,n}_\bp$ and antisymmetric imaginary part $-\Omega^{\alf\beta,n}_\bp/2$, that is, the quantum (or Fubini-Study) metric and the Berry curvature, respectively \cite{Provost1980},
%
%\begin{linenomath*}
\begin{alignat}{3}
    \label{eqn:sumgnm}
    \sum_{m\neq n} g^{\alf\beta,nm}_\bp=& &&\sum_{m\neq n} g^{\alf\beta,mn}_\bp&&=g^{\alf\beta,n}_\bp \, ,\\
    \label{eqn:sumOnm}
    \sum_{m\neq n} \Omega^{\alf\beta,nm}_\bp=&\,-&&\sum_{m\neq n} \Omega^{\alf\beta,mn}_\bp&&=\Omega^{\alf\beta,n}_\bp \, .
\end{alignat}
%\end{linenomath*}
%

\section{Charge neutrality points in Dirac semimetals} \label{sec:Semimetals}

\begin{table*}
    \centering
    \setlength{\arrayrulewidth}{0.2mm}
    \begin{tabular}{c c c c c c c c c c }
        \hline \hline
         & $\sigma^{xx}_\text{intra}$ & $\sigma^{xx,s}_\text{inter}$ & $\sigma^{yy}_\text{intra}$ & $\sigma^{yy,s}_\text{inter}$ & $\sigma^{zz}_\text{intra}$ &
         $\sigma^{zz,s}_\text{inter}$ & $\sigma^{xx,s}_\text{inter}/\sigma^{xx}_\text{intra}$ & $\sigma^{yy,s}_\text{inter}/\sigma^{yy}_\text{intra}$ & $\sigma^{zz,s}_\text{inter}/\sigma^{zz}_\text{intra}$\\[1mm]\hline
        $\lam^{(I)}_\bp$ & $\frac{1}{2}\frac{a}{b}\frac{1}{\pi}$ & $\frac{1}{2}\frac{a}{b}\frac{1}{\pi}$ & $\frac{1}{2}\frac{b}{a}\frac{1}{\pi}$ & $\frac{1}{2}\frac{b}{a}\frac{1}{\pi}$ & - & - & $1$ & $1$ & - \\[2mm]
        $\lam^{(II)}_\bp$ & $\frac{1}{\pi}$ & $\frac{a^2}{a^2+b^2}\frac{1}{\pi}$ & $\frac{1}{\pi}$ & $\frac{a^2}{a^2+b^2}\frac{1}{\pi}$ & - & - &  $\frac{a^2}{a^2+b^2}$ & $\frac{a^2}{a^2+b^2}$ & - \\[2mm]
        $\lam^{(III)}_\bp$ & {\scriptsize $ 0.197\frac{1}{\sqrt{\Gamma}}$} & {\scriptsize $0.098\frac{1}{\sqrt{\Gamma}}$ } &  {\scriptsize $0.324\sqrt{\Gamma}$ } & {\scriptsize$0.216\sqrt{\Gamma}$ }& - & - & $\frac{1}{2}$ & $\frac{2}{3}$ & -\\[2mm]
        $\lam^{(IV)}_\bp$ & $ \frac{1}{6}\frac{1}{(bc)^2}\frac{1}{\pi}$ & $\frac{1}{3}\frac{1}{(bc)^2}\frac{1}{\pi}$ & $\frac{1}{6}\frac{1}{(ac)^2}\frac{1}{\pi}$ & $\frac{1}{3}\frac{1}{(ac)^2}\frac{1}{\pi}$ & $\frac{1}{6}\frac{1}{(ab)^2}\frac{1}{\pi}$ & $\frac{1}{3}\frac{1}{(ab)^2}\frac{1}{\pi}$ & $2$ & $2$ & $2$\\[1mm]\hline \hline
    \end{tabular}
    %\internallinenumbers
    \caption{The intraband and the quantum metric contribution to the longitudinal conductivity in $x$ and $y$ directions for different models of semimetals \eqref{eqn:lam1}-\eqref{eqn:lam4} at the charge neutrality point $\mu=0$. The intraband and the quantum metric contribution are of the same order in $\Gamma$ in all cases. However, their ratio is non-universal. The numerical prefactors for $\lam^{(III)}_\bp$ are rounded for simplicity. The conductivities are given in units of $e^2/h$.}
    \label{tab:semimetals}
\end{table*}

We calculate the longitudinal conductivity explicitly for four different models of (higher-order) Dirac semimetals  
%
%\begin{linenomath*}
\begin{align}
    \lam^{(I)}_\bp&=a\,p_x\,\sigma_x+ b\,\,p_y\,\sigma_y \, ,\label{eqn:lam1}\\
    \lam^{(II)}_\bp&=\frac{a}{2}\,(p_y^2-p_x^2)\,\sigma_x+\,a\,p_x\,p_y\,\sigma_y+\frac{b}{2}\,|\bm{p}|^2\,\sigma_z\, ,\label{eqn:lam2}\\
    \lam^{(III)}_\bp&=p_x\,\sigma_x + p^2_y\,\sigma_y\, ,\label{eqn:lam3}\\[1mm]
    \lam^{(IV)}_\bp&=a\,p_x\,\sigma_x+ b\,\,p_y\,\sigma_y+c\,\,p_z\,\sigma_z \, , \label{eqn:lam4}
\end{align}
%\end{linenomath*}
%
with parameters $a,b,c>0$. $\sigma_i$ are Pauli matrices. The Hamiltonian \eqref{eqn:lam1} and \eqref{eqn:lam4} describe two- and three-dimensional Dirac cones. A quadratic band touching and a mixed linear and quadratic band touching in two dimensions are described by \eqref{eqn:lam2} and \eqref{eqn:lam3}, respectively. We fix the chemical potential $\mu$ to the charge neutrality point, so that the Fermi surface reduces to a single momentum point. The considerations for a metal do not apply since the quantum metric diverges at that momentum, so that a different scaling behavior can be expected. We perform the momentum integration over $\mathds{R}^d$ and at zero temperature. We summarize the results in Table.~\ref{tab:semimetals}.

\onecolumngrid

\section{Flat band model}
\label{sec:flatband}
As a toy model for topological flat bands we employ a square bipartite lattice where one band is flattened by longer-ranged hopping~\cite{Hofmann2020}. It reads explicitly,
%
%\begin{linenomath*}
\begin{align}
    \lam_\bp/t=&-\frac{1-\sqrt{2}}{2}\bigg[\cos\Big(2(p_x+p_y)\Big)+\cos\Big(2(p_x-p_y)\Big)\bigg]\,\mathds{1}-2\sqrt{2}\text{sin}\big(p_x\big)\text{sin}\big(p_y\big)\,\sigma_z\nonumber\\[1mm]
    &-\sqrt{2}\big[\text{cos}\big(p_y\big)+\text{cos}\big(p_x\big)\big]\,\sigma_x
    -\sqrt{2}\big[\text{cos}\big(p_y\big)-\text{cos}\big(p_x\big)\big]\,\sigma_y.
\end{align}
%\end{linenomath*}
The eigenvalues are
%\begin{linenomath*}
\begin{align}
    E^\pm_\bp=-(1-\sqrt{2})\cos 2p_x \cos 2p_y\pm\sqrt{6+2\cos 2p_x \cos 2p_y}.
\end{align}
%\end{linenomath*}
%
Numerically, we find for the energies where the bands terminate the values $E^+_\text{max}/t=3.243$,
$E^+_\text{min}/t=1.586$,
$E^-_\text{max}/t=-2.414$, and
$E^-_\text{min}/t=-2.450$. 
We place the chemical potential in the middle of the flat band, at
$\mu=(E^-_\text{max}+E^-_\text{min})/2=-2.432$. 
The quantum metric reads
%
%\begin{linenomath*}
\begin{align}
    &g^{xx,\pm}_\bp=\frac{2\big(\sin p_y\big)^2\big[1+\big(\cos p_x\big)^2\big(\cos p_y\big)^2\big]+\big(\cos p_y\big)^2\big(\sin p_x\big)^2}{\big(3+\cos 2p_x\cos 2p_y\big)^2}  \, ,\\
    &g^{yy,\pm}_\bp=\frac{2\big(\sin p_x\big)^2\big[1+\big(\cos p_x\big)^2\big(\cos p_y\big)^2\big]+\big(\cos p_x\big)^2\big(\sin p_y\big)^2}{\big(3+\cos 2p_x\cos 2p_y\big)^2} \, .
\end{align}
%\end{linenomath*}
%
We have $v^{x,\pm}(p_x,p_y)=v^{y,\pm}(-p_y,p_x)$, $v^{y,\pm}(p_x,p_y)=-v^{x,\pm}(-p_y,p_x)$, and $g^{xx,\pm}(p_x,p_y)=g^{yy,\pm}(-p_y,p_x)$. Thus, it follows that $\sigma^{xx}_\text{intra}=\sigma^{yy}_\text{intra}$ and $\sigma^{xx,s}_\text{inter}=\sigma^{yy,s}_\text{inter}$.

\section{Rhombohedral trilayer graphene}
\label{sec:RTGappendix}
The six-band model for trilayer graphene is adapted from Ref.~\cite{Zhou2021}. It holds for each valley degree of freedom and both spin degrees of freedom, meaning that the total number of bands is 24. At low electron or hole density, the system spontaneously breaks the fourfold degenerate ground state symmetry and resides in a single flavor-polarized state.
Using $\pi_\bp=\xi p_x+i p_y$ with $\xi=\pm 1$ for the valley degree of freedom, it is (choosing $\xi=1$),
%
%\begin{linenomath*}
\begin{align}
    \lam_\bp=\begin{pmatrix}
      \Delta_1+\Delta_2+\delta & \gamma_2/2 & v_0 \pi^\dag_\bp & v_4 \pi^\dag_\bp & v_3 \pi_\bp & 0 \\[1mm]
      \gamma_2/2 & \Delta_2-\Delta_1+\delta & 0 & v_3 \pi^\dag_\bp & v_4 \pi_\bp & v_0 \pi_\bp \\[1mm]
      v_0 \pi_\bp & 0 & \Delta_1+\Delta_2 & \gamma_1 & v_4 \pi^\dag_\bp & 0 \\[1mm]
      v_4 \pi_\bp & v_3 \pi_\bp & \gamma_1 & -2\Delta_2 & v_0 \pi^\dag_\bp & v_4 \pi^\dag_\bp \\[1mm]
      v_3 \pi^\dag_\bp & v_4 \pi^\dag_\bp & v_4 \pi_\bp & v_0 \pi_\bp & -2\Delta_2 & \gamma_1 \\[1mm]
      0 & v_0 \pi^\dag_\bp & 0 & v_4 \pi_\bp & \gamma_1 & \Delta_2-\Delta_1
    \end{pmatrix} \, ,
\end{align}
%\end{linenomath*}
%
where we used the shorthand $v_i=\sqrt{3} a_0 \gamma_i/2$. The parameters take the following values~\cite{Zhou2021}: $\gamma_0 = 3.1\,\text{eV}$, $\gamma_1=0.38\,\text{eV}$, $\gamma_2=-0.015\,\text{eV}$, $\gamma_3=-0.29\,\text{eV}$, $\gamma_4=-0.141\,\text{eV}$, $\delta=-0.0105\,\text{eV}$, and $\Delta_2=-0.0023\,\text{eV}$.

\twocolumngrid

% \bibliography{literature}

\end{document}